\documentstyle[aps,twocolumn]{revtex}
\input epsf

\begin{document}
\title{Coherent transients in optical lattices}
\author{O. Morsch\cite{byline},}
\address{Clarendon Laboratory, Parks Road, Oxford, OX1 3PU, United Kingdom }

\author{P.H. Jones, and D.R. Meacher}
\address{Department of Physics and Astronomy, University College London, Gower St, London WC1E 6BT, United Kingdom }
\date{\today}

\maketitle

\begin{abstract}
We use a method based on optical coherent transients to study the vibrational coherence lifetimes of
atoms trapped in the potential wells of a near-resonant optical lattice in the oscillating regime. The dependence of the
positions and widths of the vibrational Raman resonances of the coherent transient spectra on the
intensity, detuning, and geometry of the lattice beams is investigated and the results are compared
with theoretical predictions. It is shown that the principal source of broadening of the vibrational Raman resonances is the anharmonicity of the light-shift potential wells. We also perform time-resolved measurements of the vibrational excitation of atoms equilibrating in an optical lattice.
\noindent

\pacs{32.80.Pj, 42.50.Md, 39.30.+w}

\end{abstract}

\section{Introduction}
In optical lattices~\cite{meacher98b}, atoms are trapped in a periodic array of micron-sized
potential wells. Atoms confined in these potential wells exhibit a rich dynamical behaviour which
can be probed by a number of techniques, including probe absorption spectroscopy~\cite{Grynberg93a},
heterodyne fluorescence spectroscopy~\cite{Jessen92a}, recoil-induced resonances~\cite{Kozuma95a},
intensity correlation spectroscopy~\cite{Jurczak96a,Westbrook97a}, and observation of the
redistribution of photons between the lattice-inducing beams~\cite{raithel98b}. Using these
techniques, the quantized nature of the motion of the atoms in the potential wells has been
demonstrated, yielding information on the vibrational frequencies of the atoms, the motion of
wavepackets, and on the decay of coherences between the bound oscillator states. Recently, a novel
method has been suggested and experimentally demonstrated which effectively probes the coherences
set up between different vibrational states of the trapped atoms~\cite{Triche97a,Triche97b}. This
method has a number of advantages over other schemes, as we shall demonstrate. Most importantly, the
possibility of selectively exciting one particular dynamical mode of the lattice allows one to fit
theoretical predictions to experimental spectra easily and reliably.\\ In this paper we are
predominantly concerned with resonances of the coherent transient spectra corresponding to
stimulated Raman transitions between adjacent vibrational levels. The mechanisms responsible for the decay of the vibrational coherences and the resulting widths of the Raman peaks are still a subject of debate, and
a comprehensive study is as yet absent from the literature. In what follows, we
present the results of our experiments conducted on a three-dimensional lattice of cesium atoms in the {\em oscillating regime} in which the oscillatory motion of the trapped atoms is only rarely interrupted by inelastic photon
scattering events. The converse case of optical lattices operating in the jumping regime has been treated by Mennerat-Robilliard and co-workers \cite{jumping}.
In our experiments the intensity, detuning, and geometry of the lattice were varied independently, and the resulting
variations of the positions and widths of the vibrational Raman peaks were measured and compared
with theoretical models.\\
In another set of experiments, we loaded an optical lattice with atoms having an initial kinetic temperature higher than the steady-state temperature in the lattice
and made time-resolved measurements of the vibrational frequency as the atoms came to equilibrium in the lattice.
As the distribution of atoms amongst the
vibrational states evolved,  the measured vibrational frequency was observed to
change owing to the perturbations to the energy spacings of the bound levels caused by the
anharmonicity of the potential.\\ This paper is organised as
follows. In section~\ref{setup} the experimental setup is described, and in section~\ref{results} we
present the results of the experiments. Section~\ref{theory} contains a brief discussion of the
various mechanisms involved in determining the frequency widths of the vibrational Raman resonances
observed in optical lattices and a comparison of the experimental data with theoretical models.
Finally, section~\ref{conclusion} summarizes our findings.
\begin{figure}
\begin{center}\mbox{\epsfxsize 3.5in \epsfbox{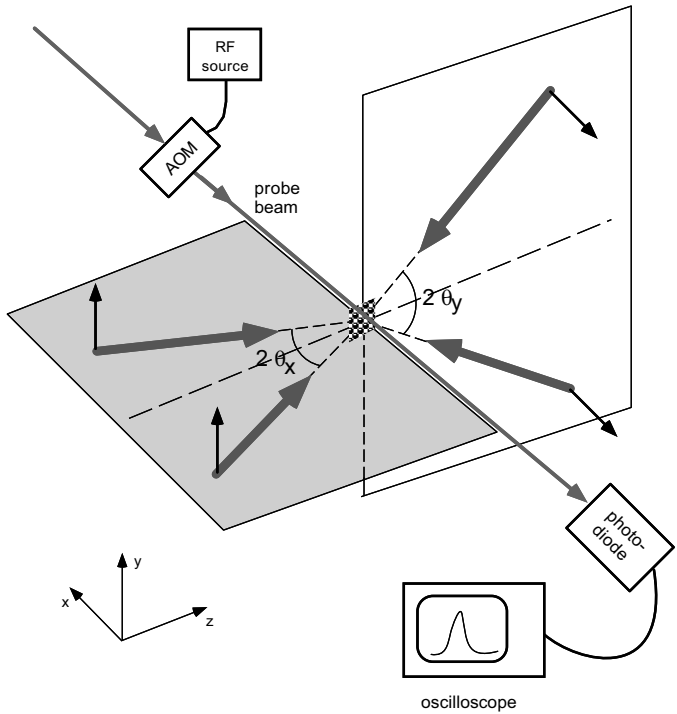}}
\end{center}
\caption{Experimental setup for a coherent transient measurement.}
\label{Fig:setup}
\end{figure}
\section{Experimental setup and procedure}\label{setup}
In our experiment, we trap and cool cesium atoms in a standard six-beam magneto-optical trap (MOT) operating on the
$6S_{1/2} (F=4)\to 6P_{3/2} (F'=5)$ transition~\cite{Raab87a}. After a $1\mathrm{s}$ trapping
period, the magnetic field gradient is switched off and the atoms are further cooled for a few
milliseconds in an optical molasses to a temperature of around $10\mathrm{\mu K}$. The molasses
beams are then switched off and the lattice beams switched on. The geometry of the lattice beams is
the standard tetrahedron configuration \cite{Petsas94a,Verkerk94a} with splitting half-angles
$\theta_x$ and $\theta_y$, where in our experiment $\theta_y$ is fixed and $\theta_x$ is variable
(see Fig.~\ref{Fig:setup}). The lattice operates on the same transition as the MOT, detuned to the
low frequency side by $3\Gamma$ to $17\Gamma$, where $1/\Gamma$ is the excited state lifetime
($\Gamma=2\pi\times5.22\mathrm{MHz}$ for cesium). The intensity of the lattice beams in our
experiment lies between $1.5\mathrm{mWcm^{-2}}$ and $3.8\mathrm{mWcm^{-2}}$. In addition to the
four lattice beams, a weak ($< 0.1\mathrm{mWcm^{-2}}$) probe beam propagates either along the $x$-
or the $z$-axis. The detuning $\delta$ of the probe laser from the lattice laser can be controlled
by an acousto-optic modulator (AOM) connected to a tunable radio-frequency source.\\ Coherent transient
spectra are recorded as follows: After the atoms have equilibrated in the lattice for a few
milliseconds, the probe beam is switched on for $20-1000\mathrm{\mu s}$ at detuning $\delta$ (see
also section~\ref{theory}). After that time, at $t=0$ the detuning is increased by
$0.9\mathrm{MHz}$ within less than $1\mathrm{\mu s}$.
After the increase in the detuning of the probe light, it no longer
interacts with the atoms and serves as a local oscillator for the heterodyne detection of the light
scattered by the atoms in the lattice.
The signal arising from the beating of the
lattice light scattered by the atoms and the local oscillator  is then recorded with a fast photodiode. Assuming that the coherently scattered light from
the lattice decays exponentially with time at a rate $\Gamma_{coh}$, the observed signal (expressed
as a voltage $V(t)$) is of the form
\begin{eqnarray}
V(t) \propto \sqrt{I_{probe}I_{coh}}\exp(-\Gamma_{coh}t)\cos[(\Delta_{jump}-\delta)t]\nonumber \\
+\frac{1}{2}(I_{probe}+I_{coh})\quad,
\end{eqnarray}
where $I_{probe}$ and $I_{coh}$ are the intensities of the probe and the coherently scattered light,
respectively, and $\Delta_{jump}$ is the amount by which the probe frequency is suddenly switched at
$t=0$. A Fast-Fourier-Transform (FFT) is then performed on the signal to yield its frequency
spectrum. Our spectra are shifted by subtracting the frequency jump
$\Delta_{jump}$, so that a Rayleigh peak (for which $\delta=0$ initially) appears at zero frequency
in our spectra.\\ In the case of the time-resolved equilibration experiment described in
section~\ref{equili}, the lattice is loaded directly from the MOT with a temperature of order
$60\mathrm{\mu K}$, and the atoms are then allowed to equilibrate in the lattice for a variable
length of time. After this period the vibrational excitation of the trapped atoms is probed and
read-out as described above.
\begin{figure}
\begin{center}\mbox{\epsfxsize 3.5in \epsfbox{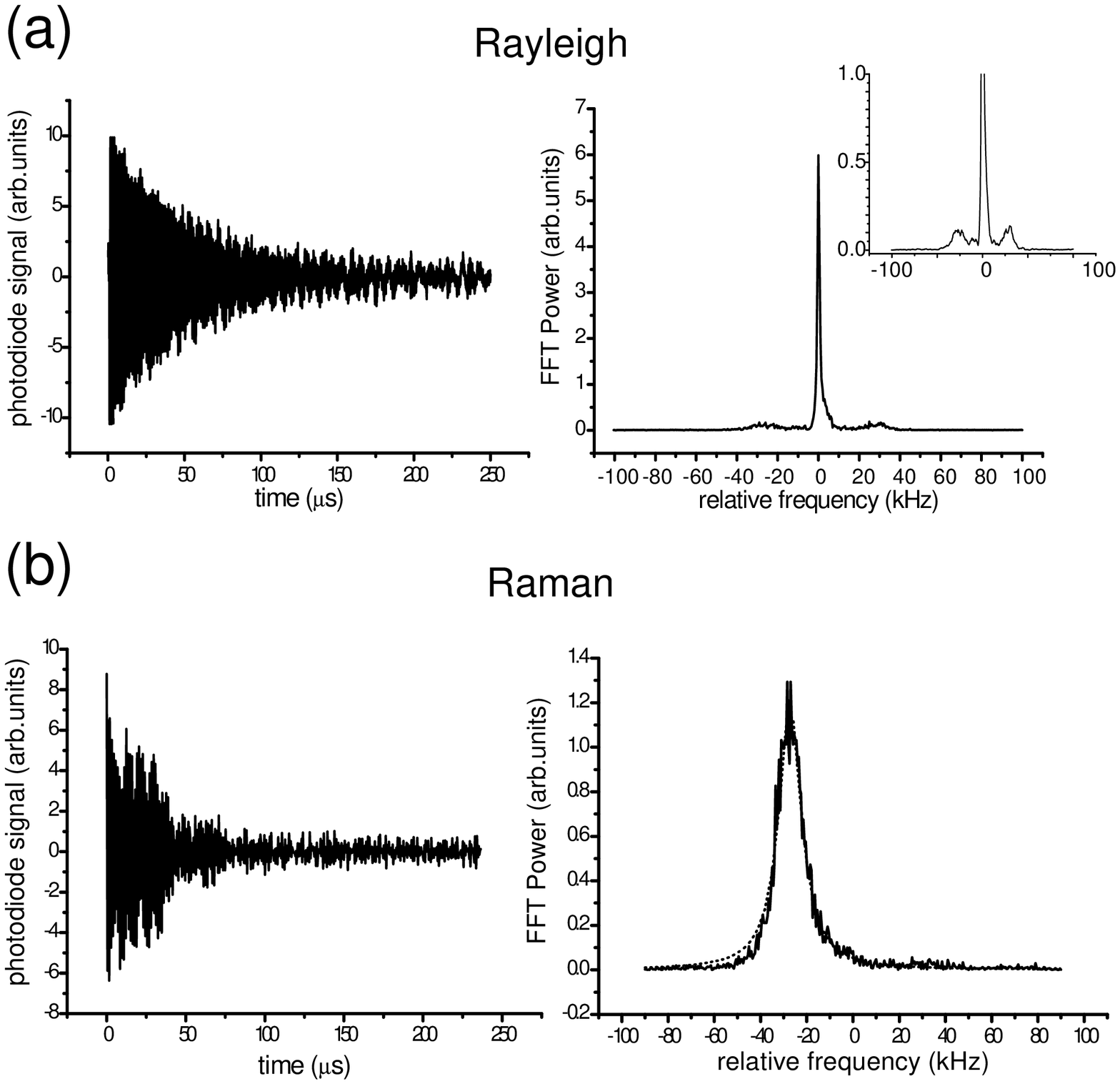}}
\end{center}
\caption{Typical experimental spectra of a Rayleigh (a) and a
Raman (b) resonance obtained by the coherent transient method. In
each case, a single feature of the complete spectrum is
selectively excited. The magnified plot inset in (a) shows more
clearly the response around the frequencies of the vibrational
Raman resonances.} \label{Fig:spectra}
\end{figure}
\section{Experimental results}\label{results}
Some aspects of the dynamical behaviour of atoms in optical lattices are better understood than
others. The vibrational frequencies of the atoms, for instance, can be rather accurately predicted
and modelled using a simple harmonic oscillator picture (for potential wells that are deep enough for
a tight-binding model to be valid)~\cite{Petsas94a}. Discrepancies with experiments can be accounted
for by the anharmonicity of the potential, which can be treated perturbatively~\cite{Gatzke97a}.\\
On the other hand, the lifetimes of atoms in these oscillator states and of the coherences between
different states, and hence the widths they give rise to in the coherent transient spectra, are not
yet well understood. Several mechanisms have been proposed that could be responsible for the
observed widths of these vibrational Raman peaks~\cite{Courtois92a,Grynberg96a,Phillips97a}. In
particular, the photon scattering rate of atoms in a light field turns out to be much larger than
the observed widths of the Raman peaks, indicating that the decay of coherences between vibrational
levels is strongly suppressed. The mechanisms responsible for this suppression will be dealt with in
more detail in section~\ref{mechs}.\\ In order to carry out a comprehensive study, we independently
varied the detuning, intensity, and geometry (characterized by the two splitting angles) of the
lattice and determined both the position and the width of the Raman peaks in the coherent transient
spectrum by fitting Lorentzians to the experimentally obtained spectra. Typical experimental results are shown in Fig.~\ref{Fig:spectra}. The signal-to-noise ratio even of a single-shot
measurement is good, so that little averaging is necessary. The signal recorded by the photodiode
is, in general, a sum of exponentially damped beat-notes between the light scattered by the trapped
atoms and the probe light. When the probe is initially tuned to the same frequency as the lattice
beams, one sees a narrow Rayleigh peak and two peaks resulting from off-resonantly excited
vibrational coherences. When detuning the probe from the lattice beams by an amount equal to the
separation between a Raman peak and the Rayleigh peak (i.e. the vibrational frequency of the atoms),
one can selectively excite the vibrational Raman coherences alone and hence obtain a spectrum that
only reflects one excitation mode of the lattice, in contrast to probe absorption spectra, where
resonances associated with many modes are superimposed. It is found that the resulting spectrum can
be well fitted to a Lorentzian, and in what follows we will refer to the width (FWHM) of such a
Lorentzian fit as the width of the Raman peak. A further advantage of the method of coherent
transients is the fact that the preparation and the read-out of the vibrational coherence grating
are separated in time. The frequency-selectivity in the excitation phase therefore depends on the
frequency bandwidth of the initial probe pulse, whilst the resolution of the spectrum only depends
on the read-out time. By making the read-out time as short as possible whilst maintaining a
reasonable resolution, time-dependent studies of the evolution of atoms in an optical lattice can be
performed. This is discussed in more detail in section~\ref{equili}.
\\ As can be seen in Fig.~\ref{Fig:spectra}, the Rayleigh peaks are of much smaller width than the Raman peaks. These widths
yield information on the rate of transfer of atoms between neighbouring wells and are discussed in
detail in~\cite{Courtois92a,Triche97b}. In this paper, however, we shall concentrate our attention
on the vibrational Raman peaks.\\
\begin{figure}
\begin{center}\mbox{\epsfxsize 3.5in \epsfbox{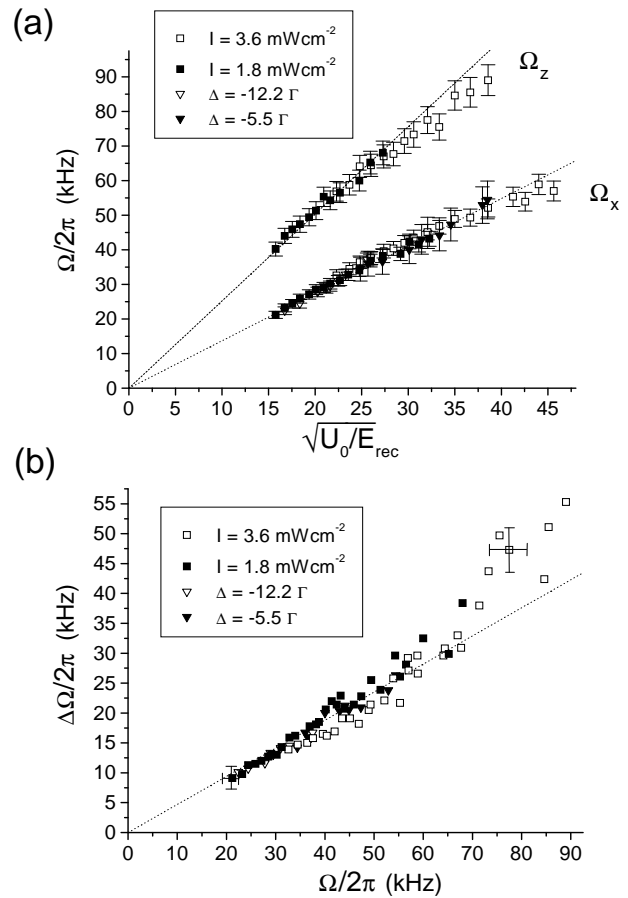}}
\end{center}
\caption{Plot of the the vibrational frequencies (a) and the Raman
widths (b). The potential well-depth was varied by changing either
the intensity or the detuning of the lattice beams, keeping the
other parameter constant. The widths in (b) are plotted against
the vibrational frequencies. For clarity, in (b) only two
representative points have error bars.} \label{Fig:intdet}
\end{figure}
\subsection{Varying intensity and detuning}\label{intdet}
For a tetrahedral lattice (see Fig.~\ref{Fig:setup}) in which the polarizations of the four lattice
beams are perpendicular to the planes spanned by each pair of nearly co-propagating beams, one can
derive simple expressions for the vibrational frequencies in a harmonic approximation of the
potential close to a minimum~\cite{Petsas94a,Gatzke97a}:
\begin{equation}\label{eqn:omxy}
\Omega_{x,y} = \omega_r \sqrt{\frac{88}{45}} \sin \theta_{x,y} \sqrt{\frac{U_0}{E_r}}
\end{equation}
and
\begin{equation}\label{eqn:omz}
\Omega_z = \omega_r (\cos \theta_x +\cos \theta_y) \sqrt{\frac{U_0}{E_r}}
\end{equation}
where  $E_r = \hbar \omega_r$ is the recoil energy with $\omega_r = 2 \pi \times 2.07\mathrm{kHz}$ for
 an optical lattice employing cesium atoms and light tuned close to the D2 resonance line
and $U_0$ is the potential well-depth given by
\begin{equation}
U_0=\hbar\Delta\frac{4 I/I_{sat}}{4(\Delta/\Gamma)^2+1}
\end{equation}
($I$ is the intensity of a single beam, $I_{sat}=1.1\mathrm{mWcm^{-2}}$, and
$\Delta=\omega_{laser}-\omega_0$ is the detuning of the lattice beams). These theoretical
predictions have been verified to yield the correct dependence on $\Delta$, $I$, and $\theta$, but
the experimental results usually fall short of the theoretical values by 15-25 percent. In our
experiment, we varied the potential well-depth $U_0$ (ranging from $U_0\approx250E_{r}$ to
$U_0\approx2100E_{r}$) both by varying the intensity (between $I=1.5\mathrm{mWcm^{-2}}$ and
$I=3.6\mathrm{mWcm^{-2}}$) and the detuning (between $\Delta=-3\Gamma$ and $\Delta=-17\Gamma$)
and measured the positions of the Raman peaks of the corresponding coherent transient spectra along
the $x$- and the $z$- directions. The results of these measurements are summarized in
Fig.~\ref{Fig:intdet} (a) and confirm the dependence of the vibrational frequencies on $U_0$ as
given by equations~\ref{eqn:omxy} and~\ref{eqn:omz}. The splitting half-angles in this experiment
were $\theta_x=41\pm1$ degrees and $\theta_y=38\pm1$ degrees, and the ratio of the slopes of
Fig.~\ref{Fig:intdet} (a) is $1.83\pm0.05$, which is close to the theoretically predicted value of
$1.67\pm0.06$.\\ The results of  measurements of the Raman widths for the cases of the probe propagating
along both the $x$- and the $z$-directions are shown in Fig.~\ref{intdet}(b) where, for clarity, most of the
error bars have been omitted~\cite{footnote3}. For the most part, the width of the Raman
peaks and the corresponding vibrational frequencies are proportional to each other, with
$\Delta\Omega\approx 0.45
\Omega$. Only for frequencies above $\approx 70\mathrm{kHz}$ do the Raman widths deviate
appreciably from this simple relationship.
\begin{figure}
\begin{center}\mbox{\epsfxsize 3.5in \epsfbox{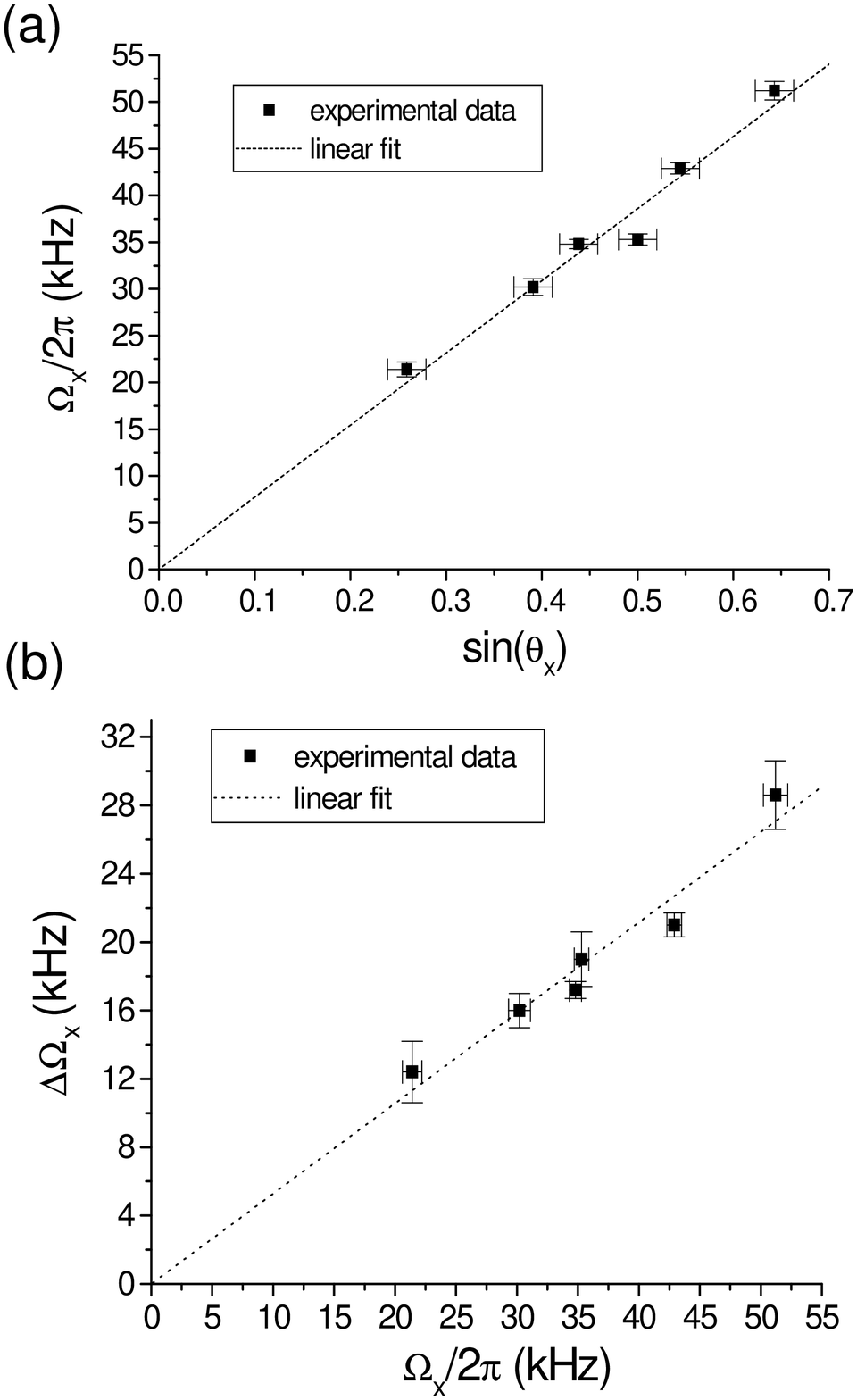}}
\end{center}
\caption{Vibrational frequencies (a) and Raman widths (b) for
different lattice geometries. The detuning and intensity of
 the lattice beams are constant. In Fig.(b), the linear dependence of the Raman widths on the vibrational frequencies is clearly visible.}
\label{Fig:geoposw}
\end{figure}
\subsection{Varying the geometry}\label{geovar}
In order to investigate the dependence of the position and width of the Raman peaks on the geometry
of the lattice, the intensity and detuning of the lattice beams (and hence $U_0$) were kept fixed
fixed at $I=3.6\mathrm{mWcm^{-2}}$ and $\Delta=-7\Gamma$, respectively, and the splitting-angle
$\theta_x$ was varied. This resulted in a change in the lattice constants in the $x$- and
$z$-directions. In our setup, $\theta_x$ could be varied between $15\pm1$ and $40\pm1$ degrees,
which according to equations ~\ref{eqn:omxy} and~\ref{eqn:omz} should result in a large variation in
$\Omega_{x,y}$, but only a small variation in $\Omega_z$. Figure~\ref{Fig:geoposw} (a) shows the
results of the measurements of $\Omega_x$ as a function of $\sin\theta_x$. As expected, there is a
linear relationship, and again the experimentally measured frequencies lie about 20 percent below
the theoretical values.\\ The measurements of the Raman widths as a function of geometry are
summarized in Fig.~\ref{Fig:geoposw} (b), where $\Delta\Omega_x$ is plotted against $\Omega_x$. As
before, there is a linear relationship between the two, with $\Delta\Omega_x\approx0.53\Omega_x$.
Within the experimental errors, this agrees with the empirical constant of proportionality found in
section~\ref{intdet}.
\begin{figure}
\begin{center}\mbox{\epsfxsize 3.5in \epsfbox{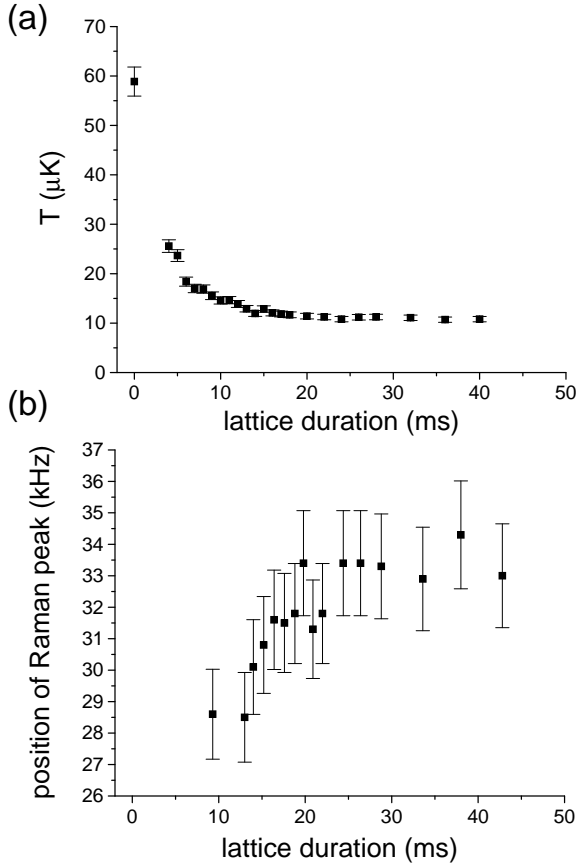}}
\end{center}
\caption{Equilibration of atoms in an optical lattice loaded
directly from a MOT with $T\approx 60\mathrm{\mu K}$, and the
change with time in the temperature (a) and observed vibrational
frequency (b) of the atoms is recorded.} \label{Fig:equil}
\end{figure}
\subsection{Time-resolved probing of the lattice evolution}\label{equili}
\begin{figure}
\begin{center}\mbox{\epsfxsize 3.5in \epsfbox{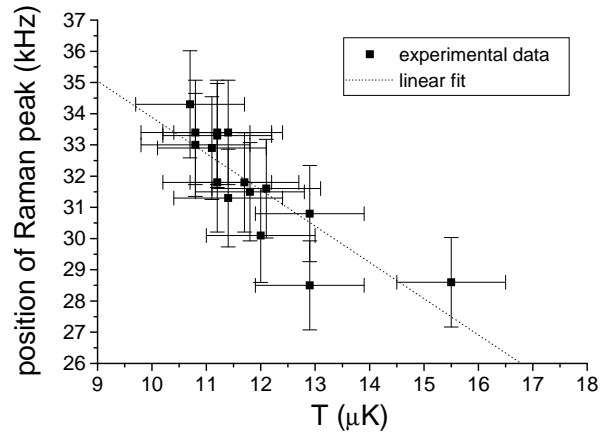}}
\end{center}
\caption{Variation of the observed vibrational frequency of the
atoms with temperature. The dotted line is a linear fit to the
data.} \label{Fig:postemp}
\end{figure}

In most of the work previously carried out on the excitation modes of atoms in optical lattices, the
atoms were allowed to reach equilibrium in the potential wells before measurements were performed.
To demonstrate that we can use the coherent transient technique for time-resolved, high-resolution
spectroscopic studies, we loaded our lattice directly from the MOT, skipping the molasses phase. The
temperature of the atoms released from the MOT was about $60\mathrm{\mu K}$. The lattice was then
switched on for a variable amount of time, after which either the temperature was measured (using
the time-of-flight method~\cite{salomon90a}), or a coherent transient spectrum was recorded in about
$1\mathrm{ms}$. Figure~\ref{Fig:equil} shows the results of the corresponding temperature and
vibrational frequency measurements. The temperature of the atoms reaches its steady-state value
governed by the lattice parameters in a few milliseconds~\cite{footnote2}. At the same time, the
measured vibrational frequency of the atoms increases as the atoms are cooled and then approaches as
steady-state value. These results can alternatively be visualized by plotting the measured
vibrational frequency against temperature (Fig.~\ref{Fig:postemp}). Evidently, the colder the atoms
are, the higher is the observed vibrational frequency. This can be understood in terms of the
anharmonicity of the lattice potential, as will be explained in more detail in
section~\ref{anharmo}. Briefly, the hotter the atoms are, the larger the population of higher-lying
(and more anharmonically shifted) vibrational states in the potential wells will be, which results
in a lower observed vibrational frequency.
\section{Theoretical discussion}\label{theory}
\subsection{Mechanisms contributing to the width of the vibrational Raman peaks}\label{mechs}
As mentioned before, whereas the dependence of the vibrational frequencies on the lattice parameters
is well understood and can be theoretically modelled, the widths of the Raman peaks of the spectra
have so far resisted a comprehensive explanation. In this section, we shall summarize the models
proposed so far and discuss them in the light of our findings. Also, we give a brief overview of the
results obtained by other groups and compare them to our own experimental results.\\ The first rate
to consider when dealing with the lifetimes of vibrational levels in optical lattices and coherences
between those levels is the photon scattering rate $\Gamma'=\Gamma U_0/\hbar\Delta$. After the first
observation of distinct Raman resonances in probe-absorption spectra it was soon realized that the
lifetimes could not be determined by $\Gamma'$ since that would have meant that (for most
experimental parameters) the resonances would not have been resolvable. The experimentally observed
widths, however, were more than an order of magnitude smaller than those predicted by the photon
scattering rate. Courtois {\em et al.}~\cite{Courtois92a} proposed the Lamb-Dicke
effect~\cite{dicke53a} arising from sub-wavelength localization of the atoms as an explanation for
the unexpected narrowness of the resonances~\cite{footnote1}. According to this theory, the rate of
inelastic scattering events in which the vibrational quantum number changes is reduced by the
Lamb-Dicke factor $(n+1)E_r/\hbar\Omega$ in favour of elastic scattering events involving no change
in vibrational level. For typical parameters, the Lamb-Dicke factor is of the order of $1/10$ or
less, so that the inelastic scattering rate $\Gamma''\propto\Gamma'(n+1) E_r/\hbar\Omega$ comes
close to the experimentally observed widths. In experiments involving shallow potentials with small
Lamb-Dicke parameters, however, Kozuma {\em et al.} still found lifetimes that were much longer than
predicted by the Lamb-Dicke model. Phillips {\em et al.} interpreted this as evidence for the
existence of the transfer in inelastic scattering processes of coherence between different pairs of
vibrational levels. Also, results of other experiments suggested that the broadening of the
vibrational Raman resonances was dominated by the anharmonicity of the lattice potential, which was
further demonstrated through the observation of revivals in wavepacket oscillations. In the
following, we will show that our experiments support that interpretation. Before discussing the
experimental findings in detail, however, we shall give a brief outline of the effects of
anharmonicity in optical lattices.
\subsection{Effects of anharmonicity}\label{anharmo}
\begin{figure}
\begin{center}\mbox{\epsfxsize 3.5in \epsfbox{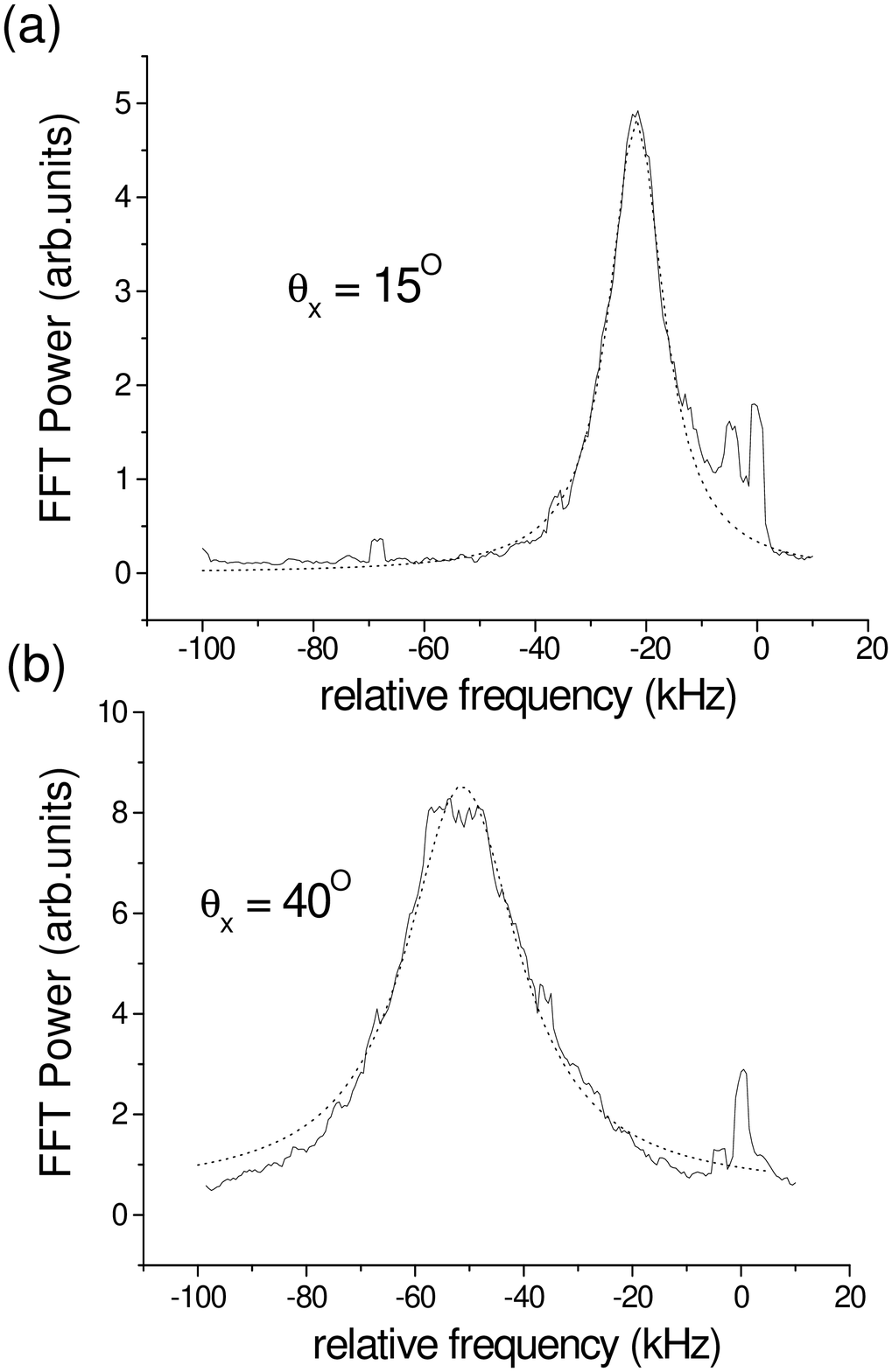}}
\end{center}
\caption{Example of Raman peaks for two different geometries of
the lattice with identical detuning and intensity of the lattice
beams. When the splitting-angle is increased, both the position
and the width of the Raman peak change. The dotted lines are
Lorentzian fits.} \label{Fig:geoex}
\end{figure}

When deriving the expressions for the vibrational frequencies of atoms in optical lattices
(equations~\ref{eqn:omxy} and~\ref{eqn:omz}), the true light-shift potentials are often approximated
by harmonic potentials characterized by the curvature of the lattice potentials close to a local
minimum. The effect of the anharmonicity of the potential can be treated perturbatively by
considering the correction due to the fourth-order
 term in the expansion of the potential in first-order perturbation theory. Taking
into account the geometrical factor $\sin\theta_x$ characterizing the lattice constant in the
$x$-direction, this leads to the vibrational frequency $\Omega_{n,n+1}$ associated with a pair of
adjacent vibrational levels characterized by $n$ and $n+1$ being reduced with respect to the
harmonic frequency $\Omega_{0}$ so that
\begin{equation}\label{anhshifts}
\Omega_{n, n+1}\approx\Omega_{0}-\omega_r (n+1)\sin^2\theta_x\quad.
\end{equation}
Assuming that in an optical lattice the vibrational levels are populated according to a Boltzmann
distribution~\cite{castin92a}, and that the ratio $C=k_B T/2U_0$ is a constant (as has been shown
experimentally~\cite{Gatzke97a}), the effect of the anharmonicity on the experimentally observed
Raman resonances can be estimated. Firstly, the observed vibrational frequency can be defined as an
average over all the anharmonically shifted pairs of vibrational levels, weighted by their
respective Boltzmann factors, leading to the following expression~\cite{Gatzke97a}:
\begin{equation}
\Omega\approx\Omega_{0}(1-2C+...)\quad.
\end{equation}
This indicates that all the vibrational frequencies should be reduced by a constant fraction, and
that that fraction will be proportional to the temperature $T$ of the atoms in the lattice for a
constant well-depth $U_0$. This satisfactorily accounts for our observations presented in
section~\ref{equili}.\\ Likewise, the broadening of the resonance due to the presence of several
vibrational frequencies can be modelled by assuming that only levels with a Boltzmann factor
(relative to that for $n=0$) larger than some small, arbitrary number $b$ contribute appreciably
to the observed spectrum. The most shifted vibrational frequency, which is a measure of the frequency
spread, is then given by
\begin{equation}
\Delta\Omega\approx\sin\theta_xC\ln(b^{-1})\omega_r\sqrt{\frac{U_0}{E_r}}=\frac{C}{2}\ln(b^{-1})\Omega_{0}\quad.
\end{equation}
This is proportional to the vibrational frequency $\Omega_{0}$ for any value of $b$,
so that we would expect that the true linewidth, calculated from a properly weighted sum over all
vibrational levels, also to be proportional to $\Omega_{0}$. This is precisely
what we observe in our experiments. Figure~\ref{Fig:geoex} illustrates the dependence of the
vibrational frequency and width of the Raman peak on the geometrical factor $\sin\theta_x$. In a
more quantitative treatment, we have performed numerical simulations of the coherent transient
spectra taking into account the frequency spread by superposing the contributions of all possible
pairs of vibrational levels. The results exhibit the same characteristics as the qualitative models
described above.\\ In principle, the anharmonicity of the potential should also manifest itself in
an asymmetry of the Raman peak. One expects a slower drop-off of the Raman peak on the low frequency
side because the anharmonic shifts are all in the same direction (see equation~\ref{anhshifts}).
This effect, however, will only be noticeable if the frequency spread due to anharmonicity is
appreciable compared to the widths of the individual contributions, which are caused by the
dissipative mechanisms described above. In our experiments, we have not found any clear evidence of
an asymmetry of the Raman peaks. In Fig.~\ref{Fig:geoex}, however, the Lorentzian fit to the broader
Raman peak (b) has a larger $\chi^2$-value then the narrower peak (a), which suggests that the
former is distorted by anharmonicity.
\subsection{Comparison of theory and experiment}
One of the main objectives of our experiments was to determine the dominant mechanism involved in
determining the width of the Raman resonances in the coherent transient spectra. In particular, we
were interested in establishing whether the Lamb-Dicke model correctly predicted the dependence of
the widths on all the experimental parameters. Since our experiment was performed on a 3-D lattice,
whereas the Lamb-Dicke treatment of Courtois~{\em et al.} is based on calculations in one dimension,
we have generalized that derivation to take into account the dependence of the lattice constant on
the geometry of the lattice. This leads to the appearance of an extra geometrical factor
$\sin\theta_x$ in the expression for the rate $\Gamma''(n)$ of inelastic Raman processes leading to a
change in the external state of the atom along the $x$-direction from an initial state with vibrational
quantum number, $n$, which then becomes
\begin{equation}
\Gamma''(n)\propto\Gamma'\sin\theta_x(n+1)\sqrt{\frac{E_r}{U_0}}\quad.
\end{equation}
This reproduces the observed linear dependence of $\Delta\Omega$ on $\Omega$ when the
splitting-angle $\theta_x$ is varied (see section~\ref{geovar}). It also turns out that when the
lattice intensity is varied, the expression for $\Gamma''$ reduces to $\Gamma''\propto\Omega$. When
$U_0$ is varied by changing the detuning, however, the relationship between $\Omega$ and $\Gamma''$
(and hence $\Delta\Omega$) is expected to be of the form $\Gamma''\propto\Omega^3$ (for large
detuning), which is not at all what we observe. In fact, upon closer inspection the values of
$\Delta\Omega$ for identical $\Omega$ but different detuning do differ slightly from each other, the
value of $\Delta\Omega$ for the larger detuning being smaller, as expected. The difference, however,
is very small and does not reflect the predicted $\Omega^3$-dependence. This is perhaps only to be expected because,
for the case of the largest detunings used in these experiments, the proximity of the neighbouring
$F=4 \rightarrow F'=4$ transition means that our approximation of an isolated transition is of dubious validity.
Nevertheless, we find that the
observed dependence of $\Delta\Omega$ on $\Omega$ agrees very well with broadening due to
anharmonicity, as derived above. These observations suggest that the main broadening mechanism is,
indeed, the anharmonicity of the potential, and the dissipative mechanisms based on inelastic photon
scattering are strongly suppressed. In the harmonic limit, where the homogeneous width is small compared to the anharmonic frequency difference between successive vibrational levels,
the suppression of line-broadening because of the transfer of vibrational coherence~\cite{Cohentannoudji92a} between pairs of levels in spontaneous
Raman processes is expected to be appreciable.
 The lifetimes of the vibrational coherences
are then determined by the cooling rate, rather than by the inelastic photon scattering
rate, which in the oscillating regime is given by $\Gamma_{cool}=4\Gamma E_r/\hbar\Delta$ ~\cite{Phillips97a} and,
in our experiments varies between $3\mathrm{kHz}$ for
small detuning to less than $1\mathrm{kHz}$ for the largest detuning employed.
For large detuning, the associated width is given by
$\Delta\Omega\propto\Omega^2$.  Conversely, in the anharmonic limit, it is the spatial
localization that is expected to give rise to the suppression of inelastic Raman processes. Clearly, the interplay between
the transfer of coherence, the Lamb-Dicke narrowing and the line broadening mechanisms is hard to model quantitatively for an intermediate case.
Nevertheless, our results strongly suggest that the broadening associated with the anharmonicity generally dominates the widths of the Raman peaks, with dissipative
mechanisms contributing at a much reduced level.

\section{Conclusion}\label{conclusion}
We used the method of coherent transients in order to measure the vibrational frequencies of atoms
in optical lattices and the widths associated with the corresponding Raman peaks of the coherent
transient spectra. Varying the intensity, detuning, and geometry of the lattices beam, we confirmed
the validity of the theoretical predictions for the vibrational frequencies based on a harmonic
oscillator model of the potential wells. Our measurements of the Raman widths as a function of
intensity, detuning, and geometry all yielded a linear dependence on the vibrational frequencies,
which strongly suggests that anharmonicity is the main broadening mechanism. This implies that the
decay of the coherence between adjacent vibrational levels is suppressed, which we ascribe to the combined effects of the
transfer of coherence between pairs of vibrational levels in inelastic Raman processes and the suppression of the inelastic rate arising from the sub-optical wavelength spatial localization.
The anharmonicity was also found to be responsible for a
reduction in the observed vibrational frequencies relative to the predictions of a harmonic model,
which was tested by measuring the vibrational frequencies of ensembles of atoms with different
temperatures in otherwise identical optical lattices.\\

\section*{Acknowledgements}
We would like to thank D. Lucas, G. Birkl, and M. Oberthaler for stimulating discussions and D.
Smith for help with the electronics. This work was supported by the  EPSRC (UK). O.M. gratefully
acknowledges support by the Studienstiftung des Deutschen Volkes.


\begin{thebibliography}{ourpaper}
\bibitem{footnote3}The errors on $\Omega$ and $\Delta\Omega$ stem from the uncertainty in the
fitting of the Lorentzian to the spectra. The smaller the detuning, the greater the uncertainty due
to the noise on the signal. The uncertainty in the intensity measurements was about 10 percent.
\bibitem{footnote2}As in~\cite{Gatzke97a}, we found a linear dependence of $T$ on the potential well-depth
$U_0$.
\end{thebibliography}
\end{document}